\documentclass[fleqn,usenatbib]{mnras}
\usepackage[utf8]{inputenc}
\usepackage[subpreambles=true]{standalone}

\usepackage{import} 

\usepackage{graphicx}
\graphicspath{ {./images/} }

\usepackage{xcolor}

\usepackage{amsmath}

\usepackage{amssymb}

\usepackage{comment}

\title[Shape noise and dispersion in PWL]{Shape noise and dispersion in precision weak lensing}
\author[Pol Gurri et al.]{
Pol Gurri$^{1}$\thanks{E-mail: pgurriperez@swin.edu.au}
Edward N. Taylor$^{1}$
and Christopher J. Fluke$^{1}$
\\
$^{1}$Centre for Astrophysics \& Supercomputing, Swinburne University of Technology, Victoria 3122, Australia.
}

\begin{document}

\maketitle

\begin{abstract}
We analyse the first measurements from precision weak lensing (PWL): a new methodology for measuring individual galaxy-galaxy weak lensing through velocity information. Our goal is to understand the observed shear distribution from PWL, which is broader than can be explained by the statistical measurement errors.  We identify two possible sources of scatter to explain the observed distribution: a shape noise term associated with the underlying assumption of circular stable rotation, and an astrophysical signal consistent with a log-normal dispersion around the stellar-to-halo mass relation (SHMR). We have modelled the observed distribution as the combination of these two factors and quantified their most likely values given our data. For the current sample, we measure an effective shape noise of $\sigma_\gamma = 0.024 \pm 0.007$, highlighting the low noise impact of the method and positioning PWL as $\sim 10$ times more precise than conventional weak lensing. We also measure an average dispersion in shears of $\xi_\gamma = 0.53^{+0.26}_{-0.28}$\,dex over the range of $8.5 < \log M_\star < 11$. This measurement is higher than expected, which is suggestive of a relatively high dispersion in halo mass and/or profile.
\end{abstract}


\section{Introduction} 
\label{sec:intro}

Relating galaxies to their dark matter halos remains a complex issue. One avenue that has proven very successful for measuring halo masses is weak gravitational lensing (WL). WL describes subtle distortions on the images of background sources when observed on sightlines travelling close to a foreground mass distribution referred to as the lens \citep[see reviews by, e.g.][]{Bartelmann01, Hoekstra08, Hoekstra13}. In simple terms, the observational signature of WL is that the observed shapes of background galaxies appear stretched in the tangential direction and contracted in the radial direction to the lens. The amount of deformation, or shear, depends on the total lensing mass and the lens--source geometry. Since the halo mass overwhelmingly dominates the lensing mass, for a particular lens--source system WL provides a direct measurement of the halo mass.

The underlying assumption of these WL methods is that observed galaxies are randomly oriented, and thus the mean axis ratio should be 1 (i.e. circular) when averaged over large ensembles. Any deviation from that idealisation can then be attributed to the effects of lensing. While assuming that an individual galaxy is perfectly circular is clearly not true, it remains reasonable if averaged over a sufficiently large number of randomly oriented galaxies. The statistical description of errors associated with the assumption of random orientations (or combined circularity), commonly referred to as `shape noise', is by far the largest source of uncertainty in WL \citep[e.g.][]{Leauthaud07,Kuijken15}.

The common approach to reduce the effects of shape noise is to co-add or `stack' results from many individual sources at the cost of only being sensitive to an average shear signal for the ensemble \citep[but see, e.g.][]{Dvornik20, Taylor20}. However, several different studies have tried to reduce shape noise by also identifying the best morphologies and/or galaxy properties to target \citep[e.g][]{Niemi15, Croft17}, including additional morphological information \citep[e.g.][]{Brown11, Huff13} or even using machine learning approaches \citep[e.g.][]{Springer20}.

Despite all these efforts, shape noise still remains the limiting factor in WL studies, which restricts their sensitivity to stacked (or averaged) halo measurements. As an example, WL has been used to measure the SHMR, an important relation that connects galaxy's stellar masses to their halo counterparts, based on stacked lensing profiles for galaxies binned by stellar mass \citep[e.g.][]{Sifon15, vanUitert16, Dvornik20}. However, while it is expected that two halos with the same mass will harbour galaxies with different stellar mass and properties \citep[e.g.][]{Mandelbaum06, Li13}, WL remains insensitive to the particularities of the individual galaxies. As a result, WL studies are unable to provide strong constraints on, for example, the dispersion around a median SHMR \citep[but see e.g.][]{Taylor20}, a property that encodes important information to understand the different effects that dark matter has on the formation and evolution of galaxies.

Aiming to avoid the need for stacking, a new way to perform WL with the potential to be sensitive to individual galaxies was proposed by \citet{Blain02}, followed by \citet{Morales06} and \citet{deBurghDay15, deBurghDay15b}, and first applied to data by \citet[][hereafter Paper I]{Gurri20}. The new methodology, PWL, builds on the assumption that the velocity fields of stably-rotating galaxies can be fitted accurately by pure circular rotation motions \citep[e.g.][]{Mo10}. Under that assumption, the observed velocity fields of galaxies must be axisymmetric (their maximum and minimum velocity gradients must be orthogonal). As lensing shears the shape of galaxies, their observed velocity maps get distorted as well and are no longer axisymmetric. The amount of non-axisymmetry is proportional to the shear, and thus, can be related to the halo mass of the lens. Using these ideas, \citet{deBurghDay15} undertook a numerical study to understand the data requirements and limitations of PWL and predicted that PWL measurements could be achieved with current telescopes.

In Paper I, we have selected, observed and analysed 19 systems using PWL to demonstrate the usability of PWL by showing that the combined lensing signal of our sample dominated over any source of random noise: the  variance weighted mean of observed shears for the sample is $0.020 \pm 0.008$, which represents a detection of the lensing signal at $>99$\% confidence. However, the limiting precision of PWL methods is still to be determined. 

Similar to the idealisation of combined circularity in the shapes of galaxies from conventional WL, it is not expected that all galaxies rotate with perfectly circular motions. We use the term `dynamical shape noise' to refer to the statistical description of errors associated with the assumption of axisymmetry or stable circular rotation. To test the impact of dynamical shape noise, in Paper I we have analysed a set of unlensed galaxies and found that the extent to which the assumption holds is sample-specific, but showing that dynamical shape noise has the potential to be much lower than the limiting shape noise in conventional WL experiments: a simple estimate from Paper I being of order $\sim 0.03$ compared to shape noises of $\sim 0.2$ -- $0.3$ from conventional WL studies (see Section \ref{sec:conclusion} for more discussion on this topic).

Here we analyse the PWL measurements presented in Paper I, with the aim of describing a methodology to determine the limiting precision of PWL for specific samples. At the same time, we will use this methodology to disentangle the amount of noise in PWL from a real astrophysical signal motivated by the expected dispersion in the SHMR.

We have structured this paper as follows: In Section \ref{sec:data} we briefly present the sample and results from Paper I which will be used for the study. In Section \ref{sec:analysis} we provide an analysis of the possible sources of scatter between expected shear values and measured ones. These include a description of how deviations from axisymmetry propagate through PWL measurements and the distribution of shears we expect from a dispersion in the SHMR.  In Section \ref{sec:results} we present our measured constraints on shape noise and the dispersion together with a discussion on the implications of the findings. We offer a brief summary and vision for the future in Section \ref{sec:conclusion}.

\section{DATA}
\label{sec:data}

In this paper, we focus on 21 PWL shear measurements for a sample of 19 weakly lensed galaxies. Here we only briefly review the sample, which is fully described in Paper I. 

Our targets (or sources) have been selected from a compendium of spectroscopic redshifts surveys, including the 2dF Galaxy Redshift Survey \citep[2dFGRS;][]{Colless01}, 6dF Galaxy Survey \citep[6dFGS;][]{Jones09}, Sloan Digital Sky Survey \citep[SDSS;][]{Aihara11}, and the Galaxy And Mass Assembly survey \citep[GAMA;][]{Driver11,Liske15}. We have identified source candidates as being in close projection to an intervening lens galaxy, with large enough redshift separation ($\Delta z > 0.05$) and small enough angular separation ($\sim$ 10s of kpc) to ensure an appreciable degree of lensing.

For each candidate lens--source pair, we obtain a median expectation for the shear as follows. We use optical/near infrared photometry from GAMA--KiDS/VIKING \citep{Wright16}, SDSS, or Pan-STARRS1 \citep{Chambers16} to derive a stellar mass estimate for the lens galaxy, following \citet{Taylor11}. We then use the \citet{vanUitert16} SHMR determination to obtain a median expectation for the halo mass of the lens based on its stellar mass. A median expectation for the shear, $\gamma_{pred}$ then follows via:
\begin{equation}
    \gamma(r) = \frac{\overline{\Sigma}(r) - \Sigma(r)}{\Sigma_{crit}}
\end{equation}
where $\Sigma(r)$ is the surface density at $r$, $\overline{\Sigma}(r)$ the mean surface density inside the given radius $r$, defined as $\overline{\Sigma}(r) = 1/r \int\Sigma(r')dr'$ and $\Sigma_{crit}$ the critical surface density which depends (only) on the lens--source geometry \citep[e.g.][]{Miralda-Escude91a,Wright00}. Following \citet{vanUitert16}, we have assumed a \citep[][NFW]{Navarro96} halo profile  and a mass--concentration relation based on \citet{Duffy08}. We only considered systems with a non-negligible amount of shear, $\gamma_{pred} > 0.001$. The values for the expected shears within our sample span the range $0.001 < \gamma_{pred} < 0.012$; with a mean value of $\langle \gamma_{pred} \rangle = 0.005$.

Our source targets have been selected to be bright (apparent $i$-band magnitude $< 17.4$) and large so they can be well-resolved ($\sim 5\arcsec$). When selecting targets to observe, we tried to minimise potential errors and maximise signal by giving preference to sources with 1) spiral morphologies, for which the assumption of stable rotation is more robust (see Paper I), and 2) orientations that maximise the observable shear via PWL, where the major velocity axis of the source is at $\sim \pm 45^\circ$ with respect to the lensing direction (see Section \ref{sec:analysis}). The sources in our sample span the redshift range $0.06 < z < 0.15$, with stellar masses in the range $9.5 < \log M_\star < 11.25$.

Our only selection criterion for lenses was to discard systems where the lens was part of a cluster/rich group or had signs of a recent merger or disturbance. We placed this selection in order to ensure that we measure the effects of individual, undisturbed halos. To the extent that our target selection is based on the properties of the background source galaxies and not the lenses, our sample represents an unbiased set of 19 central lenses in the local Universe ($0.006 < z < 0.06$) and with stellar masses spanning the range $8.5 < \log M_\star < 11$. The median and mean values for $\log M_\star$ for the lenses are $10.49$ and $10.28$, respectively. 

All source galaxies were observed with the Wide-Field Spectrograph (WiFeS), an optical-slicing integral field unit (IFU) mounted on the Australian National University (ANU) 2.3m telescope \citep{Dopita07, Dopita10}. The resulting data consist of gas velocity fields with a minimum of 50 well-resolved spatial elements with useful signal to noise ($S/N > 2$) and low velocity uncertainty ($\sigma(v) < 50$\,km/s). This data quality is predicted to be sufficient to obtain an unbiased measure of the lensing signal \citep[see][]{deBurghDay15}. We obtained shear measurements, $\gamma^{obs}$, by modelling every velocity field as a linearly-lensed stably rotating disk, with the shear included as a free parameter (See Paper I for a complete description). The measured variance-weighted mean shear via PWL is $\langle \gamma^{obs} \rangle = 0.020 \pm 0.008$, which represents a detection of the lensing signal at $>99$\% confidence.

\begin{figure}
    \centering
      \includegraphics[width=\columnwidth]{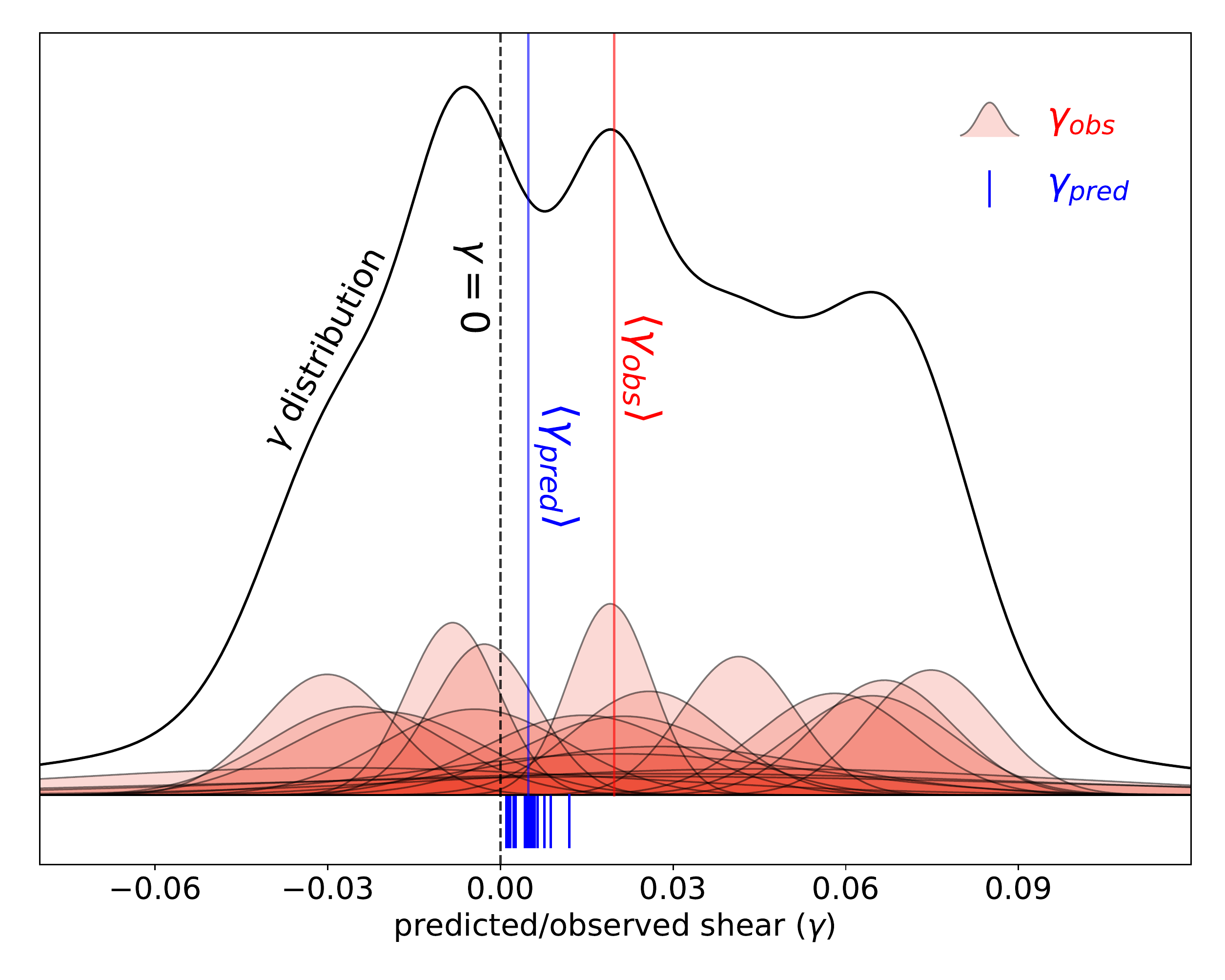}%
      \caption{ 
      We plot the total observed distribution of shears from the sample described in Sec.\ \ref{sec:data}. In red we plot the posterior probability density function (PDF) for each shear measurement (small distributions at the bottom) together with its predicted shear as a small blue line. The total shear distribution, plotted in a black solid line, represents the distribution of observed shears convolved with the measurement errors. The signature of lensing is to shift and/or skew the distribution towards positive values. The red and blue large lines represent the variance weighted mean of the measured shears and the mean of the predicted shears respectively. Negative shear measurements indicate the presence of shape noise, while the skewness of the total distribution suggests a non-negligible dispersion in shears consistent with a dispersion in the SHMR.
      }
      \label{fig:observed_distribution}
\end{figure}

\section{Analysis: signal and noise in the distribution of observed shears}
\label{sec:analysis}

Figure \ref{fig:observed_distribution} shows the distribution of observed shears across our sample.  As discussed in Paper I, the fact that the mean observed shear is positive shows that the lensing signal dominates over any source of noise ($>99$\% confidence), at least when averaged over our ensemble of 21 measurements. At the same time, the scatter of the observed values around our expectations is large: $\sim$2.5 times larger than can be explained by the formal statistical measurement errors. Our goal here is to better understand the distribution of observed values, including the contributions of potential astrophysical sources of scatter and of error/uncertainty.

There are at least three mechanisms capable of creating a significant dispersion in the observed distribution of shears: 
\begin{enumerate} 
\item random/statistical measurement errors associated with PWL measurements;
\item deviations from axisymmetry in the intrinsic velocity fields for some or all of the target galaxies (i.e.\ there is some effective shape noise);
\item real astrophysical variations in the properties of the lenses, relative to the median SHMR.
\end{enumerate}
We also note that another possibility is that our median expectations for the observed shears, which are derived using the results from \citet{vanUitert16}, are incorrect. There could be issues in the SHMR determination itself, and/or deviations from an NFW profile, and/or halo substructure. Apart from mentioning this possibility and including a short discussion in Section \ref{sec:results}, further consideration of this point is beyond the scope of this paper.

The first source of dispersion (i.e.\ statistical measurement errors/uncertainties) is explicitly accounted for in the process of fitting for the observed shears. As these measurements have been derived using Markov Chain Monte Carlo (MCMC) modelling, we have fully propagated the observational uncertainties from the reduced and calibrated spectra through to the inferred shears. These errors are shown in Figure \ref{fig:observed_distribution} as red shaded areas.

The fact that the observed distribution is broader than can be explained by the measurement errors alone
shows that the second and/or third mechanisms (i.e.\ effective shape noise and/or genuine astrophysical differences in the lenses) are significant. While it is not possible to constrain those phenomena in an individual case basis, it is possible to statistically recover information from the ensemble. In the following subsections, we focus on the distinct phenomenology of these two processes.

\subsection{Dynamical and effective shape noise}

The necessary assumption that underpins our analysis is that, apart from the action of lensing, our target galaxies are intrinsically axisymmetric, as would be expected for pure circular rotation.  In the presence of a bulge, bar, spiral arms, warp, interaction, inflow/outflow, etc., this idealisation will necessarily be wrong at some level. The critical question is how deviations from pure rotation and strict axisymmetry (what we call `dynamical shape noise') will, in a statistical sense, limit the precision of individual shear measurements. 

\begin{figure*}
    \centering
      \includegraphics[width=2\columnwidth]{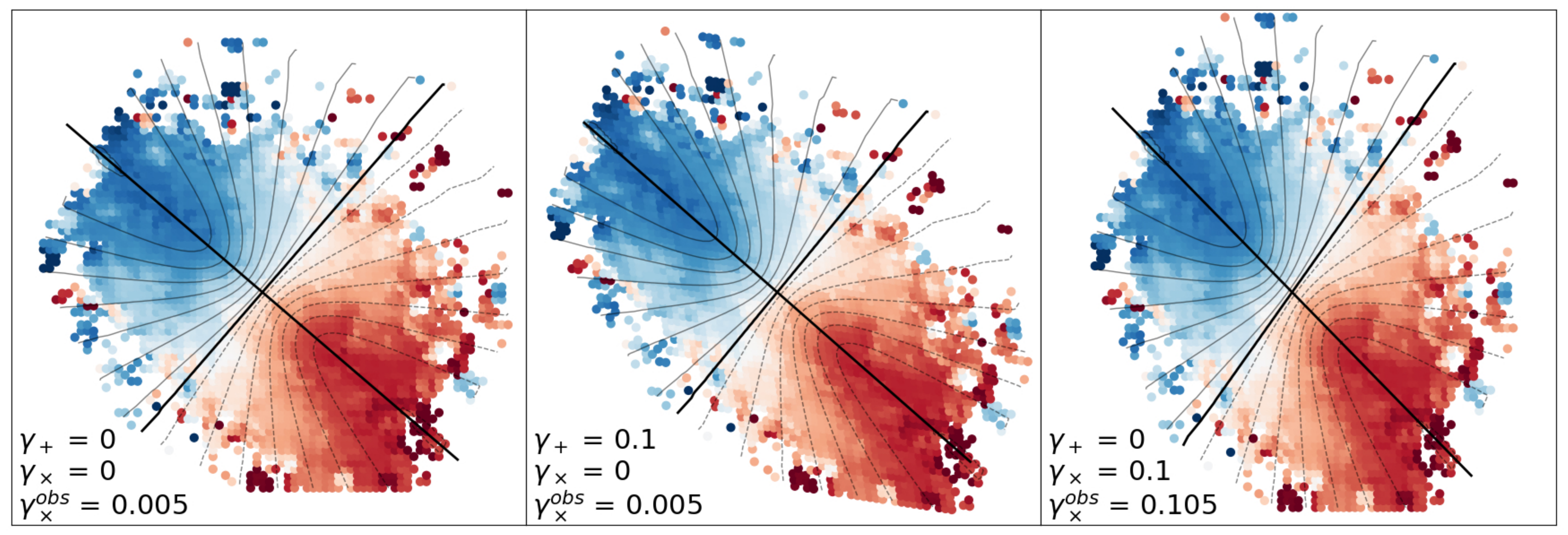}
      \caption{ Left: Velocity map of an unlensed galaxy from CALIFA (UGC00005). In grey lines, we plot the best fit model for the velocity field and in solid lines the maximum and minimum velocity gradients. This galaxy is not perfectly axisymmetric, and our best fit model returns a $\gamma^{obs}_\times = 0.005$. Middle: The same galaxy with the effects of a fictitious lens aligned with the minimum velocity gradient ($\phi = 90^\circ$). This panel highlights the effects of $\gamma_+$, which are to expand and contract the galaxy through the tangential and radial lensing directions respectively. The effects of $\gamma_+$ do not change the axisymmetry of the galaxy, and our best fit model recovers $\gamma^{obs}_\times = 0.005$ consistent with the original non-axisymmetry. Right: The same galaxy, this time with a fictitious lens aligned at $\phi = 45^\circ$ (to the left of the panel) inducing a $\gamma_\times = 0.1$. This panel highlights the effects of $\gamma_\times$, which are to change the relative angles between the major and minor velocity axis and as a result, change the axisymmetry of the galaxy. In this scenario, our model would recover a $\gamma^{obs}_\times = 0.105$ as the intrinsic axisymmetry is linearly added with the lensing effect of $\gamma_\times = 0.1$. This is the lensing signal that PWL studies are sensitive to.
      } 
      \label{fig:figure2}
\end{figure*}

To guide this discussion, in Figure \ref{fig:figure2} we use an unlensed but non-axisymmetric galaxy to illustrate the two phenomenologically distinct components of a linearised shear. The data shown have been obtained from the Calar Alto Legacy Integral Field Area \citep[CALIFA;][]{Sanchez10}. In the left panel, we show the galaxy as observed and the shear that we would infer from its non-axisymmetric velocity field. In the middle and right panels, we have recreated the same CALIFA galaxy under the different effects of linearised WL.

\begin{figure}
    \centering
      \includegraphics[width=\columnwidth]{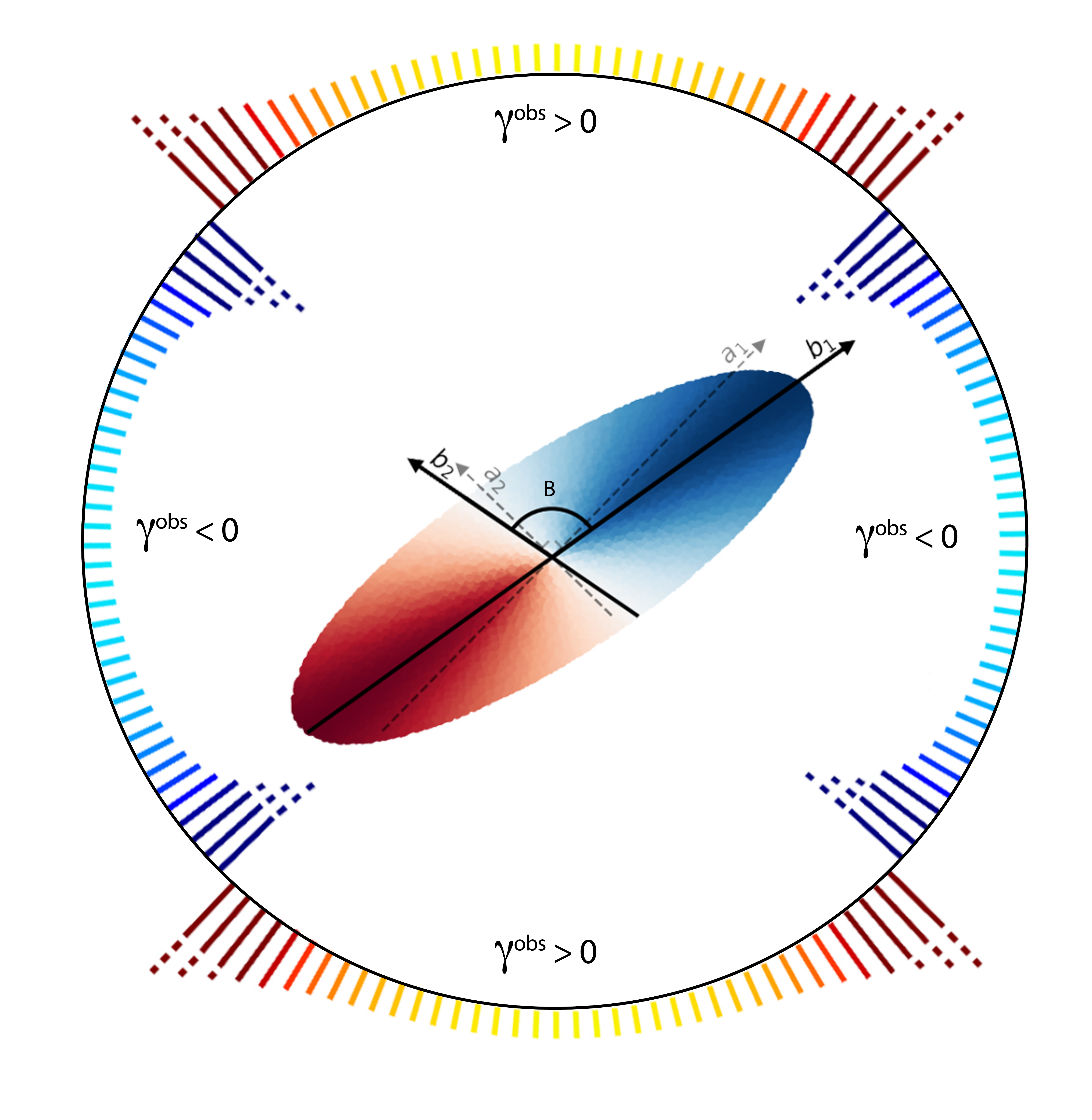}
      \caption{
      We present an example of a galaxy that is affected by dynamical shape noise and thus mimics a shear signal.  In our example, the unlensed galaxy has major and minor axis $b_1$ and $b_2$ that are not orthogonal (B$\, > 90^{\circ}$) due to some sort of intrinsic irregularity (similar to Figure \ref{fig:figure2}a). Axis $a_1$ and $a_2$ represent the axisymmetric system that PWL assumes. In this example, PWL would mistakenly measure a shear signal $\gamma^{obs}$. The sign of the recovered $\gamma^{obs}$ depends on the lensing angle $\phi$, defined as the angle between $a_1$ and the lensing direction. The outer ring with lines represents all possible directions to a lens. Lines facing outwards (red) represent directions for which we would measure a positive shear, while inwards lines (blue) show directions that would lead to a negative shear measurement with the length of each line representing the magnitude of the measured shear (dotted lines are not in scale). Because the position of galaxies is random, we are equally as likely to obtain a positive or a negative measurement of $\gamma^{obs}$. This ensures that the distribution of dynamical shape noise will be symmetrical and centred at zero. This plot also highlights the importance of targeting systems where the lensing direction is not aligned with the major or minor axis of the galaxy, as the error in the measured shear is far greater in these situations.
      }
      \label{fig:lensdirection}
\end{figure}

The first-order component of linearised WL, convergence, stretches the image of the lensed galaxy by the same amount in all directions. This magnifying effect is observationally indistinguishable from a `larger' galaxy. The second-order effect, shear ($\gamma$), comes from differential magnification in different directions. Shear is a tensor property that can be factorised into two independent and orthogonal components, $\gamma_+$ and $\gamma_\times$, which we refer to as the `plus' and `cross' terms of the shear. The relative strength of the two components is governed by the angle between the major axis of the source and the direction to the lens, $\phi$, which we refer to simply as `the lensing angle'; specifically: $\gamma_+ = |\gamma|\cos(2\phi)$ and $\gamma_\times = |\gamma|\sin(2\phi)$.

In the middle panel of Figure \ref{fig:figure2} we have positioned an imaginary lens at $\phi = \pm 90^{\circ}$ (aligned with the minor velocity axis of the galaxy) to show the pure plus term scenario; i.e.\ $\gamma_+ = \gamma$ and $\gamma_\times = 0$. The effect of $\gamma_+$ is to stretch the image of the source through one of its axes while compressing it in the other. As a result, if the direction to the lens is aligned with one of its axis ($\phi = 0^{\circ}, \pm 90^{\circ}, 180^{\circ}$), only the effects of $\gamma_+$ are present, and the changes to the observed velocity field are indistinguishable from an increase (or decrease) in the scale radius and a decrease (or increase) in the inclination angle. In other words, because the plus term does not change the degree of axisymmetry in the observed velocity field, our PWL approach is not sensitive to $\gamma_+$.

To show the effect of the cross term, in the right panel of Figure \ref{fig:figure2}, we have positioned an imaginary lens at $\phi = 45^{\circ}$ (to the left of the figure) so that $\gamma_\times = \gamma$ and $\gamma_+ = 0$. In this case, the stretching and compressing of the galaxy happens through the bi-section of the axis, which changes the angle between the major and minor axes of the galaxy and breaks the axisymmetry in the velocity field \citep[see][]{Blain02, Morales06, deBurghDay15}. We can see how $\gamma_\times$ drives a noticeable change in the axisymmetry of the velocity field as the major and minor axes form a greater angle than what they did. PWL operates by measuring this cross-component of the shear, $\gamma_\times$, through its distorting effects on the symmetry of the projected velocity field. Then, knowing the lensing angle $\phi$, the cross-term is used to obtain the total shear as $\gamma = \gamma_\times / \sin(2\phi)$.

Since PWL is only sensitive to the cross term of the shear, any error in the shear measurement arising from a non-axisymmetry in the intrinsic (unlensed) velocity field can only impact the inferred $\gamma_\times^{obs}$. We describe the error due to a non-axisymmetry, $\varepsilon_\times$, as the (generally unknown) value for $\gamma_\times$ that we would infer in the absence of any lensing (e.g. $\varepsilon_\times = 0.005$ for the galaxy in Figure \ref{fig:figure2}). This error then propagates as $\gamma_\times^{obs} = \gamma_\times^{true} + \varepsilon_\times$, with $\gamma_\times^{true}$ being the astrophysical shear due to lensing (also unknown).  This behaviour can be seen in the right panel of Figure \ref{fig:figure2} where $\varepsilon_\times$ is linearly added to the true lensing ($\gamma_{\times}^{true} = 0.1$) and the observed cross shear is the sum of the two $\gamma_\times^{obs} = 0.105$. As the quantity of interest is $\gamma$, estimated as $\gamma_\times / \sin (2\phi)$, the full propagation of errors becomes:
\begin{equation}
    \label{eq:gamatot}
    \gamma^{obs} = \gamma_{true} + \frac{\varepsilon_\times}{\sin(2\phi)} + \sigma^{obs}
\end{equation}
where $\gamma^{obs}$ is the total shear that we observe/measure, $\gamma_{true}$ is the total, unknown astrophysical shear and $\sigma^{obs}$ represents a random error/statistical uncertainty on the total shear measurement. This equation also highlights why we have preferred targets where $\sin(2\phi) \sim 1$, as the noise/error term $\varepsilon_\times$ can be seen to be amplified for each individual target by a factor of $1/\sin(2\phi)$.

\newpage

Figure \ref{fig:lensdirection} shows explicitly how a deviation from axisymmetry propagates through to the final value of $\gamma^{obs}$ depending on the lensing angle. Note that, all else being equal, the $\sin(2\phi)$ dependence means that the contribution of $\varepsilon_\times$ to $\gamma^{obs}$ will flip signs if the lensing angle is changed by $\phi \pm 90^\circ$. For an ensemble, dynamical shape noise propagates through to the final measurements according to the distribution of $\varepsilon_\times/\sin(2\phi)$. Since galaxy orientations are random, the lensing angles $\phi$ and $\phi \pm 90^\circ$ are equally likely, and therefore we expect as many positive contributions to $\gamma^{obs}$ as negative. 
A similar argument can be made based on the viewing/inclination angle: the sign of $\varepsilon_\times$ itself will flip depending on whether a galaxy is viewed from above ($i > 0$) or below ($i < 0$). As a result, {\em the distribution of $\varepsilon_\times/sin(2\phi)$ is necessarily symmetric and centred at zero}, which means that the effect of dynamical shape noise can only be random, symmetric, and zero-centred noise.

While we have framed this discussion around intrinsic irregularities in the velocity field of galaxies, 
it is worth highlighting that the arguments pertain to any and all sources of shape noise. For example, if there were some distortions in the detector astrometry, the effect would be to introduce an error in the observations via the cross term of the shear.\footnote{In Paper I, we have limited detector-tied, systematic contributions to the observed shear for our sample to be $\varepsilon_\times < 0.02$ (95\% conf. ), i.e.\ small compared to our expectations for dynamical shape noise.}
Again, the resulting error distribution would be symmetric and zero-centred, with the relevant quantity being the uniformly distributed angle between the source position angle and detector roll angle.
Similarly, any sources of error tied to the lens coordinate system (e.g. halo triaxiality) will propagate to shear measurements through to a symmetric and zero-centred error distribution, depending on the relation between $\phi$ and the position angle of the lens.

The aggregate of all effects capable of mimicking a shear define an `effective shape noise', which operates as a quantitative, statistical description of the validity of the assumption of axisymmetry in the observed velocity fields, regardless of the origin of the non-axisymmetry. For the purposes of this paper, we will describe this effective shape noise in terms of the RMS error in the inferred value of $\varepsilon_\times$, which we denote as $\sigma_\gamma$. In making this choice, we are implicitly or explicitly assuming Gaussian statistics.  While this is the simplest and most convenient choice, one could adopt more flexible parameterisation for the distribution: like the Student or $t$-distribution, with a shape parameter to describe the relative power in the wings. In principle, the exact shape of the distribution could be measured with a sufficiently large ensemble, especially if spanning a broad range in $\phi$. However, in practice, we cannot support either approach with the current data. Nevertheless, as we present next, the fact that the effective shape noise must be symmetric and zero-centred is enough to distinguish between this kind of random noise and a genuine astrophysical signal.

\subsection{Dispersion in SHMR}

The third possible source of scatter between predicted and observed shears are real astrophysical differences in the lensing mass distributions. Our median shear predictions are based on a one-to-one relation between stellar mass and shear (assuming a median SHMR and a fixed concentration as a function of halo mass). However, at a given stellar mass we expect galaxies to span a range of halo masses around the median halo prediction (i.e.\ a dispersion in the SHMR).

Since for isolated lenses shear is proportional to the halo mass 
\footnote{We note that shear is proportional to the mass contrast between a region and its surroundings, and so not strictly proportional to the mass directly. That said, given the relatively strong shears within our sample, the cosmological shear from structure around or along the line of sight is negligible: in the order of $\gamma < 10^{-4}$ compared to our expectations of $\gamma = 0.005$ to $0.01$}.
(via the excess surface density, and all else being constant), a dispersion around the median SHMR should propagate directly into a dispersion in the observed shear distribution around our median shear expectations. 

\newpage

At a fixed mass, the dispersion in the SHMR is usually described as a log-normal distribution \citep[e.g.][]{Behroozi10, Reddick13, RPuebla15, Zu15, Lange19}. While most of these studies consider a log-normal dispersion in stellar masses (at fixed halo mass), from an observational point of view the natural choice is to describe it as a log-normal dispersion in halo masses (at fixed stellar mass). While some models predict the amount of dispersion to be a function of mass, we describe it with a single value. This choice implies that if there was a mass-dispersion dependence, we would only be sensitive to an averaged dispersion across the range of our lenses. We return to this issue and its implications in Sec. \ref{sec:results}.

From these considerations, there are three significant expectations about the astrophysical shear distribution of isolated lenses: 
1.) because mass is strictly positive, the distribution should be strictly positive;
2.) the median SHMR provides a median expectation for the shear distribution for any specific lens--source pair; 
and 3.) the prospect that the dispersion around the SHMR should be approximately log-normal means that the dispersion in the astrophysical shear distribution will be skewed to higher values. 
These features of the astrophysical shear distribution are in direct contrast to the expectations for the effective shape noise which is symmetric and zero-centred.

As such, for a given system we describe the distribution of possible shears as a log-normal distribution of width $\xi_\gamma$ and median $\gamma_{pred}$. In a similar way that using a normal distribution lets us quantify the effective shape noise in terms of the RMS in shears, using a log-normal distribution lets us quantify astrophysical dispersion in terms of the RMS in log-shear. An advantage of describing astrophysical differences directly in terms of shear is that we are agnostic to the precise nature of what might be the cause of these variations and we are simply recovering the amount of variation. We return to this issue later in the discussion.

\subsection{Modelling the shear distribution}

 \begin{figure*}
    \centering
      \includegraphics[width=2\columnwidth]{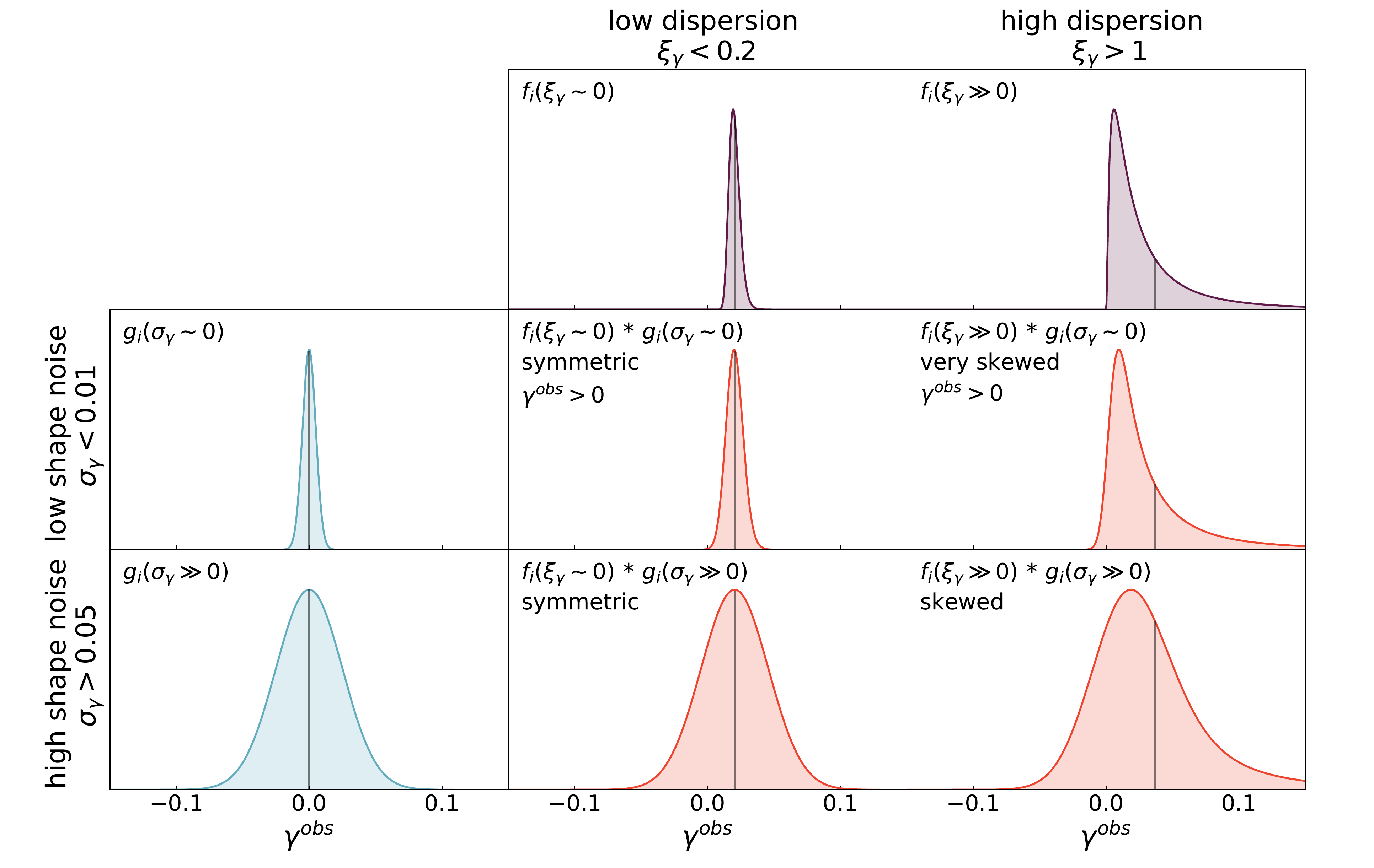}%
      \caption{We present different possible shear distributions coming from scenarios with low and high dispersion and shape noise. The top two panels (purple distributions, denoted with $f_i$) display a low and high log-normal dispersion in shears consistent with a dispersion in the SHMR. Both distributions share a median equal to $\gamma_{pred}$. The mean of each distribution is represented with a grey line. The left two panels (blue distributions, denoted with $g_i$) display the Gaussian distribution expected from a low and high effective shape noise. Both distributions share a zero mean and median. The four middle panels (red distributions, denoted with $f_i * g_i$) show the expected distribution of shears as the convolution of the corresponding dispersion and dynamical shape noise. We can see that the mean of the corresponding dispersion distribution is conserved in the convolution, and how negative values of $\gamma^{obs}$ indicate the presence of non-negligible effective shape noise. From the right column, we can observe that skewness is a good indicator of a high dispersion.
      }
      \label{fig:convolutions}
\end{figure*}

In line with the arguments above, we create a generative model for the observed shear distribution in terms of a log-normal distribution (to encapsulate an astrophysical dispersion in the properties of lenses) and two Gaussian distributions (to describe the effective shape noise and observational errors respectively). Then, the probability of observing a shear $\gamma^{obs}$ is defined as the convolution of these three distributions.

In more detail, for a given lens--source pair (denoted with the sub-index $i$), the lens mass and the lens--source geometry define $\gamma^{pred}_i$ and $\phi_i$. We can then define the probability that the true astrophysical shear takes the value $\gamma'$ as a function of the dispersion $\xi_\gamma$:
\begin{equation}
    f_i(\gamma' | \xi_\gamma ) = \frac{1}{\gamma' \xi_\gamma\sqrt{2\pi}\ln(10)} \exp\left[-{\frac{\log_{10}\left(\gamma' / \gamma^{pred}_i\right)^2}{2\xi_\gamma^2}}\right]
\end{equation}

Experimental error and effective shape noise mean that the shear that we observe, $\gamma_i^{obs}$, will be different from the true astrophysical value of $\gamma'$ by some amount. The propagation of these two sources of error is described as the convolution of two Gaussian distributions, which is analytic: the result being a Gaussian distribution with standard deviations added in quadrature. With our PWL observations determining the values of $\gamma_i^{obs}$ and $\sigma_i^{obs}$, we can define the probability of observing $\gamma_i^{obs}$ as:

\begin{equation}
    g_i(\gamma_i^{obs} - \gamma' | \sigma^{\text{tot}}_i) = \frac{1}{\sigma_i^{\text{tot}}\sqrt{2\pi}} \exp\left[-\frac{\left(\gamma_i^{obs} - \gamma'\right)^2}{2\left(\sigma_i^{\text{tot}}\right)^2}\right]
\end{equation}

\noindent where 

\begin{equation}
\left(\sigma^{\text{tot}}_i\right)^2 = \left(\frac{\sigma_\gamma}{\sin(2\phi_i)}\right)^2 + \left(\sigma^{obs}_i\right)^2 .
\end{equation}

Finally, to recover the total probability of observing a shear $\gamma_i^{obs}$ we need to account for all possible scenarios leading to the measurement of $\gamma_i^{obs}$ given the true astrophysical shear taking a value of $\gamma'$ multiplied by the probability of the true astrophysical shear taking the value of $\gamma'$. Formally, this is done by taking a probability weighted integral over all possible values of the (unknown) true astrophysical shear:
\begin{multline}
\label{eq:like}
    \text{P}_i(\gamma^{obs}_i | \xi_\gamma, \sigma_\gamma) = (f_i * g_i)(\gamma^{obs}_i) = \\
    \int^{\infty}_{0} f_i(\gamma' | \xi_\gamma)\ g_i(\gamma^{obs}_i - \gamma' | \sigma^{\text{tot}}_i)\ \delta \gamma'
\end{multline}
As this integral is not analytic it needs to be computed numerically for each pair of $\xi_\gamma, \sigma_\gamma$.

To illustrate how the observed shear distribution depends on the two parameters $\xi_\gamma$ and $\sigma_\gamma$, in Figure \ref{fig:convolutions} we plot possible shear distributions for a galaxy resulting from different scenarios with high and low dispersion and effective shape noise.  By comparing Figure \ref{fig:observed_distribution} to Figure \ref{fig:convolutions} we can make some preliminary conclusions about the dispersion and effective shape noise of our sample. First, the distribution cannot be easily explained with only noise as there is a clear skew towards positive values. This can only be explained with a large real astrophysical signal coming from a dispersion term (similar to the right column in Figure \ref{fig:convolutions}). Together with the positive mean of $\gamma^{obs}$, this further confirms that the lensing signal dominates over noise for the ensemble. At the same time, because large negative values of $\gamma^{obs}$ can only be explained by an effective shape noise, the observed negative outliers inform about a non-negligible effective shape noise (similar to the bottom row in \ref{fig:convolutions}).

\section{Results and Discussion}
\label{sec:results}

\begin{figure*}
    \centering
      \includegraphics[width=2\columnwidth]{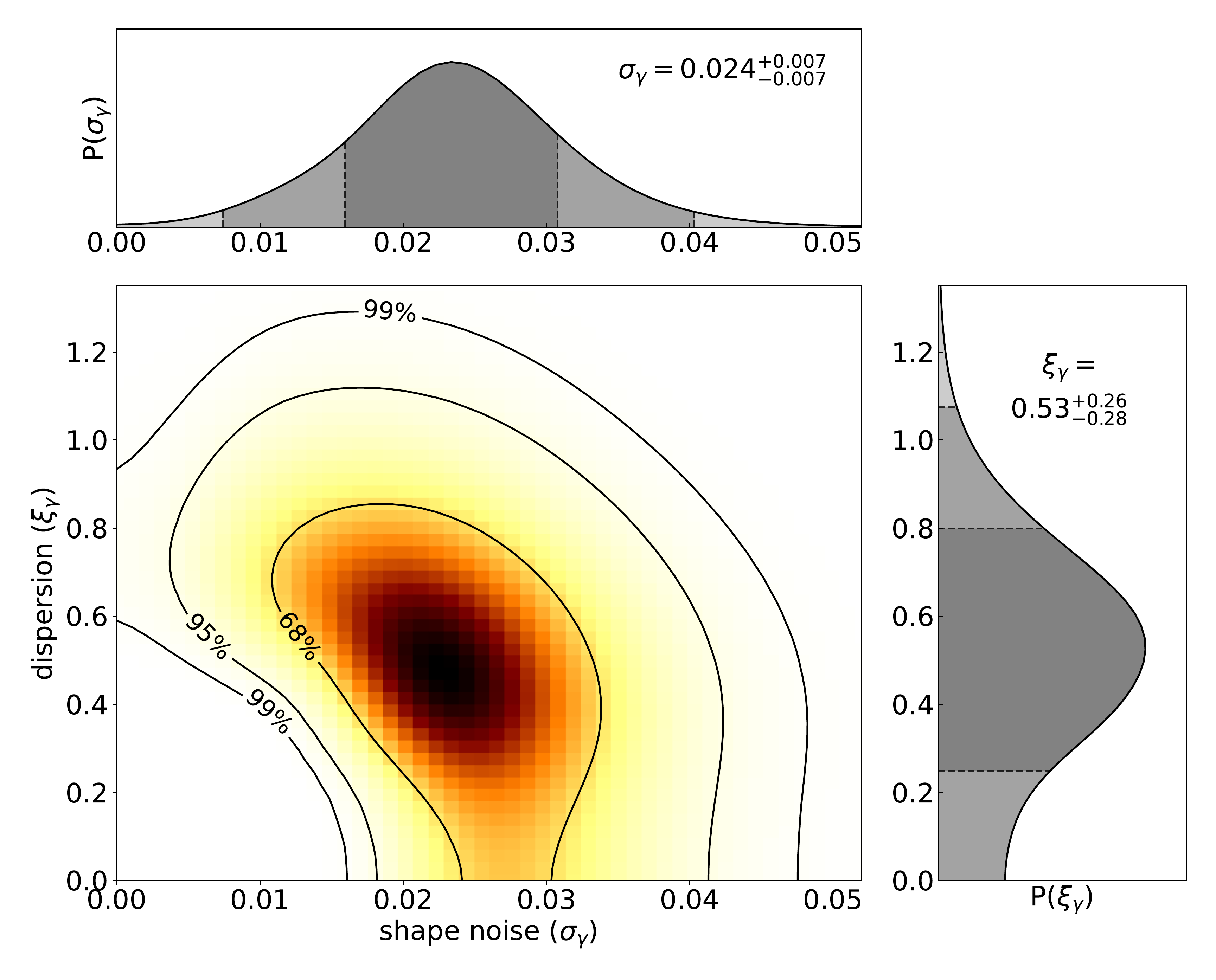}
      \caption{ Joint constraints on the dispersion and dynamical shape noise coming from the maximum likelihood grid search applied to the Paper I's sample. The colour intensity levels represent the likelihood of the data coming from that particular combination. We have added contours highlighting the 1-,2- and 3-$\sigma$ confidence levels in the measurement. On the right-hand side and on top we present the marginalised distributions of the two parameters, the dispersion and the dynamical shape noise respectively. We measure a maximum likelihood dynamical shape noise of $\sigma_\gamma = 0.024 \pm 0.007$, which is slightly lower than expected, and a higher-than-expected dispersion in shears of $\xi_\gamma = 0.53^{+0.26}_{-0.28}$\,dex.
      }
      \label{fig:dispvssn}
\end{figure*}

With a descriptive model for the observed shear distribution, the aim of this paper is to quantify the impact of an effective shape noise (which defines the limiting precision of PWL in our sample) and disentangle it from an astrophysical dispersion. Within the framework of Bayesian statistics, the probability of $\xi_\gamma$ and $\sigma_\gamma$ can be determined by analysing the joint likelihood of observing the full sample as a function of the unknown dispersion and effective shape noise, $\mathcal{L}(\xi_\gamma, \sigma_\gamma) = \sum \ln P_i(\gamma_i^{obs} | \xi_\gamma, \sigma_\gamma)$. To do so, we have assumed uniform priors on both parameters and mapped the likelihood function $\mathcal{L}$ across the $\xi_\gamma$ and $\sigma_\gamma$ parameter space.

In Figure \ref{fig:dispvssn} we present the joint constraints on the dispersion and effective shape noise for our sample. We find that the maximum likelihood values for the effective shape noise and dispersion in shears are $\sigma_\gamma = 0.024$ and $\xi_\gamma = 0.53$ dex, respectively. The marginal constraints on each parameter are also shown.
Because the constraint on the dispersion is non-Gaussian, we prefer to report our results in terms of the 16/50/84 percentiles: Our results with uncertainties become, $\sigma_\gamma = 0.024 \pm 0.007$ and $\xi_\gamma = 0.53^{+0.26}_{-0.28}$\,dex. Note that there is some covariance between the two parameters within these ranges, such that if the effective shape noise is higher, then the astrophysical dispersion will be lower, and vice versa. 

Our results point to a low effective shape noise ($\sigma_\gamma < 0.04$ at $95\%$ confidence), might suggest a higher than expected astrophysical dispersion (although uncertainties are large), and place a constraint on $\xi_\gamma > 0 $ at the $ 90\%$ confidence level. We discuss these points and their implications in the next two subsections.

\subsection{The effective shape noise is low}
\label{sec:4.1}
Our primary goal has been to describe a method to constrain the limiting precision of PWL. Because the effective shape noise is, in general, sample-specific (especially depending on how well the central assumption of stable rotation holds), the first significant result is that we have shown how the effective shape noise can be inferred directly from the main science sample. This can be viewed as an avenue to self-calibrate the amount of effective shape noise, which we anticipate will be an important component and/or a cross-check on future PWL studies. This will be especially important if, as in our case, an astrophysical parameter of interest has the potential to covary with the effective shape noise. To provide a point of reference, the described methodology would be similar to using the tangential projection of the shear to identify/limit random and systematic errors in conventional WL \citep[e.g.][]{Viola15}. 

Probably the most significant implication of our analysis is the demonstration that the impact of an effective shape noise can be very small for PWL studies. For our sample, we have measured a $\sigma_\gamma \sim 0.024$, and $\sigma_\gamma < 0.04$ (95\% conf.). While shape noise is sample-specific, our recovered value is in good agreement with our expectations from the analysis of unlensed CALIFA galaxies in Paper I, where we found that for an appropriately selected sample like ours, the effective shape noise would be $\sim 0.03$.

When comparing the impact of shape noise to that of observational uncertainties, even for such a small sample their contributions are similar;
with $\langle \sigma^{obs} \rangle = 0.024$ and $\langle 0.024/\sin(2\phi) \rangle = 0.027$, highlighting the low impact of shape noise in our PWL measurements. Similarly, when considering the mean observed shear $\langle \gamma^{obs} \rangle = 0.020 \pm 0.008$ (see Paper I), a shape noise contribution of $0.024 / \sqrt{21} \sim 0.005$ is actually smaller than the measurement errors in the mean. In other words, {\em even with only 21 measurements, our mean shear measurement is currently limited by data quality, more so than shape noise.}

Compared to conventional WL, where shape noise is $\sigma_\gamma \sim 0.2 - 0.3$ \citep[e.g][]{Leauthaud07,Niemi15, Kuijken15}, an effective shape noise of $\sigma_\gamma \sim 0.024$ means that PWL measurements are $\sim 10$ times more precise. Even when compared against proposed methods to reduce conventional shape noise, PWL is still $5-10$ less affected by noise. In other words, each PWL measurement in our sample carries similar information to $\sim 100$ {\em equally lensed} galaxies analysed through conventional WL studies. The fact that the $\sim 100$ galaxies would need to be equally lensed is very significant, pointing towards the potential for PWL to analyse rare and high-value targets where it is impossible or just very costly to build signal through sheer weight of numbers. 
We highlight however, that the greatest strength of PWL is not simply a higher signal to noise ratio, but the possibility of avoiding stacking, which opens an avenue to new kinds of measurements.

The most consequential assumption of describing the effective shape noise within our sample is that of Gaussian statistics. In Section \ref{sec:analysis} we provided strong arguments to support the modelling of the effective shape noise as a symmetric and zero-centred distribution, but the exact shape of the distribution is beyond our ability to predict. In the immediate context of this paper, it is conceivable that the true noise distribution has broad wings and/or that our sample includes one or two large and positive outliers that would lead us to infer a larger dispersion than we might otherwise. However, even if that were the case, outliers alone cannot explain the clear skew observed in Figure \ref{fig:observed_distribution}. Further, their presence would mean that real limiting effective shape noise for well-behaved targets is even smaller than we think.

These issues can in principle be addressed with more robust statistics. Large sample sizes would ensure that outliers carry less weight and lose the potential to affect results, as well as allowing us to test different distributions. A second option would be to make use of the potential for stellar dynamics to further validate the assumption of stable rotation. From our analysis of unlensed galaxies in Paper I, we found that the best test of axisymmetry is requiring consistency between stellar and gas velocity fields: the gas and stellar velocity fields giving inconsistent results is a clear indication of complex dynamics beyond pure rotation. In such a way, it would be possible to identify and exclude targets that are likely to have greater dynamical shape noise. Further, where the two measurements are consistent, the stellar-derived shear measurements acts as an independent measurement, effectively doubling the sample size.

\subsection{The astrophysical shear dispersion is high}

\begin{figure}
    \centering
      \includegraphics[width=\columnwidth]{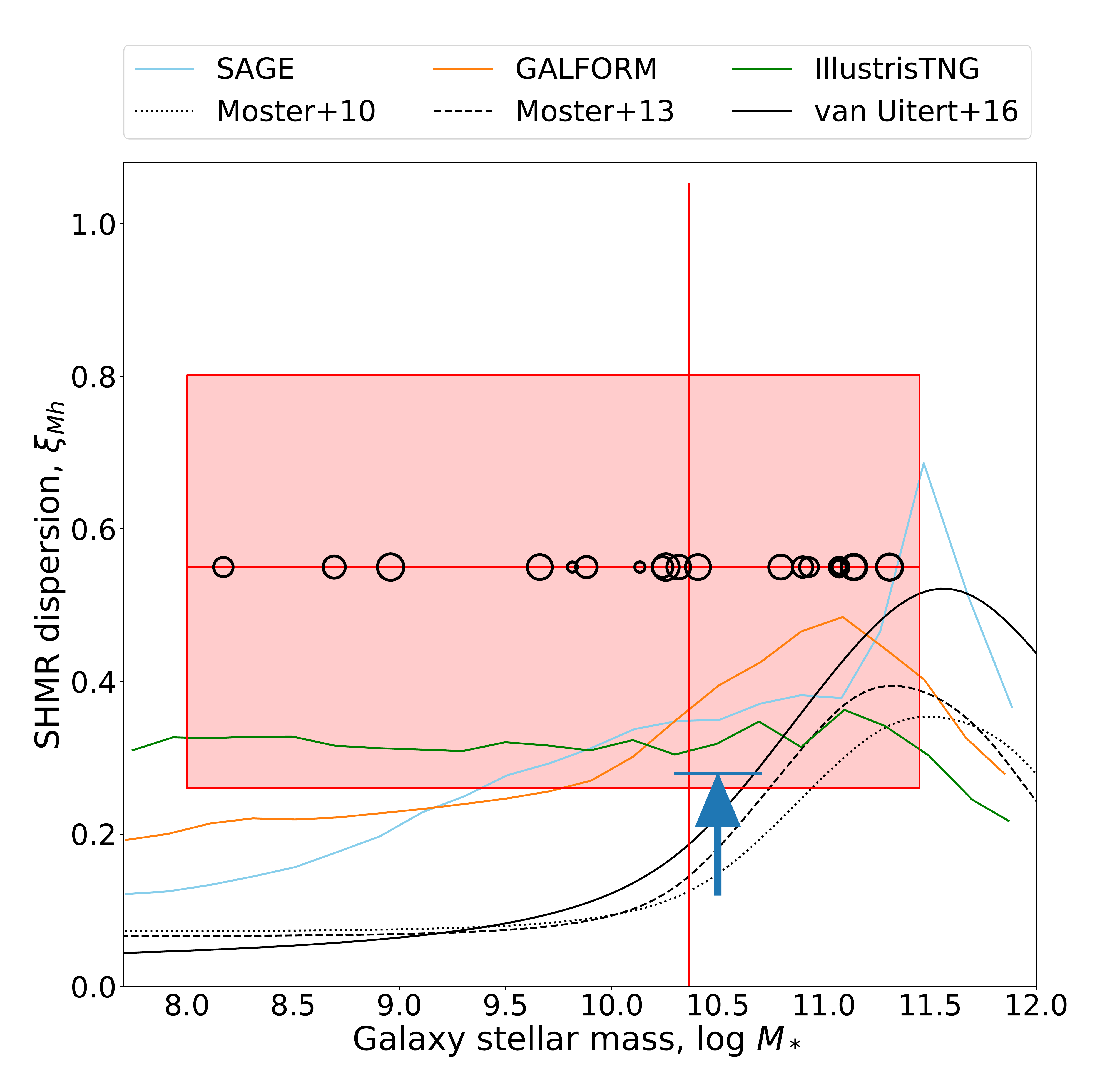}%
      \caption{
      Our measurement of the SHMR dispersion in context. With a red box and red lines we show our 84\% and 95\% confidence constraints. The circles show the masses of individual lenses, with sizes reflecting their relative weights in the measurement. Our measurement represents a weighted average dispersion across the range of $8.7 < \log(M_\star) < 11.3$. For comparison, we have plotted with a blue arrow the upper limit in the dispersion at at $\log(M_\star) \sim 10.5$ from \citet{Taylor20} obtained though lensing data alone. We plot with colored lines the predictions from three popular models from \citep[][respectively]{Stevens16, Lagos18, Pillepich18}, and with black lines the results from halo occupation modeling style analysis \citep[from][]{Moster10, Moster13, vanUitert16}, which typically assume a low value for the SHMR dispersion as a function of halo mass. Our results suggest a substantially higher dispersion than other studies.}
      \label{fig:dispcomparison}
\end{figure}

A secondary goal of this paper has been to quantify the amount of dispersion in the lensing properties of our sample of galaxies. From our analysis, we can conclude that there is a non-zero astrophysical dispersion in shears at the 90\% confidence level. In a similar way to how we argued in Paper I that the positive mean shear shows that the astrophysical lensing signal dominates over any source of noise, the non-zero constraint for the dispersion shows that the observed shear distribution for our sample of just 19 galaxies is genuinely probing real astrophysical differences in the lensing properties of galaxies at $z < 0.06$. 

A naive approach to allow for comparison would be to interpret the dispersion in the observed shear distribution as reflecting a dispersion in the SHMR under the simple assumption that shear is proportional to halo mass $\gamma \propto M_h$ so that $\xi_{M_h} \sim \xi_\gamma = 0.53^{+0.26}_{-0.28}$\,dex. While we have described the dispersion using a single number, the SHMR dispersion is itself likely to vary as a function of mass. In this case, our measured value should be interpreted in terms of a weighted average across the mass range of our set of lenses ($8.5 < \log(M_\star) < 11$). As more data becomes available, the same process can be applied to smaller ranges of $M_\star$ to measure how the dispersion changes as a function of mass.

For comparison, in Figure \ref{fig:dispcomparison} we show the results of different studies/models reporting a dispersion in halo masses. The comparable values coming from theoretical/computational models would be $0.35$\,dex, $0.37$\,dex, $0.32$\,dex \citep[][respectively]{Stevens16, Lagos18, Pillepich18}. Results from halo occupation modelling usually used in weak lensing return averaged values of $0.19$\,dex, $0.22$\,dex, and $0.27$\,dex \citep[respectively]{Moster10, Moster13, vanUitert16}. Taking our results at face value, this naively inferred $\xi_{M_h} = 0.53^{+0.26}_{-0.28}$\,dex would be rather high compared to models and especially compared to weak lensing/halo occupation results.

In this simplistic comparison, we have neglected the contribution of other expected sources of dispersion. As one example, the comparison above neglects both scatter in the halo shape/concentration at fixed halo mass and any covariance between halo mass and halo concentration at fixed stellar mass. If, following \citet{Duffy08}, we assume a log-normal dispersion in concentration of $\xi_c = 0.12$\,dex, we find that such dispersion in concentration would account for $\xi_\gamma \sim 0.1$. As a result, our measurement of the dispersion in the SHMR would be lowered to $\xi_{M_h} \sim 0.4$: still high, but in slightly better agreement with models and past observational results.

Another possibility to resolve this apparent tension is that our median expectations for the predicted shears are wrong. To explore this we have repeated our analysis with the inclusion of a multiplicative scale factor $A$ to all predicted shears. We found a maximum likelihood $A = 3.5 \pm 1$ and a measured dispersion of $\xi_\gamma = 0.2 \pm 0.2$. While these findings suggest that different SHMR parametrisations could result in tighter measurements, changing $A$ on its own would violate some of the constraints used to derive the SHMR: particularly, the halo mass function and/or stellar mass function constraints.  Interestingly, our dynamical shape noise estimator was not covariant with $A$, meaning that different SHMR parametrisations would not impact our measurement of the limiting precision of the method. Based on this, future research could investigate new constraints on the SHMR including PWL data.

In conclusion, although uncertainties are high, we measure a larger than expected dispersion in shears, but with multiple nonexclusive possibilities for what might be driving this. Without speculating, possible explanations include: the dispersion around the SHMR is much larger than suggested by previous observational studies, and/or our assumed SHMR relation is off, and/or we are seeing other sources of variation like halo substructure, variations in the inner slope, etc.

\section{Summary and Conclusions}
\label{sec:conclusion}

In this paper, we have presented a methodology for PWL studies to statistically measure the limiting precision of a given sample and differentiate the contribution of noise terms from that of an astrophysical signal.

We have used the sample from \citet{Gurri20} which consists of 21 individual shear measurements from the velocity fields of 19 weakly lensed galaxies. The source galaxies are bright ($15 < i_{mag} < 17$) and nearby ($0.06 < z_S < 0.15$) to ensure well-resolved velocity fields. The lenses in the sample represent an unbiased selection with masses within the range of $8.5 < \log(M_\star) < 11$ and redshift range $0.006 < z < 0.06$. We have used \citet{vanUitert16} SHMR determination to determine halo masses for the lenses and predict a median shear signal expected for each system. Expected shears span the range $0.001 < \gamma_{pred} < 0.012$ and have a mean value of $\langle \gamma_{pred} \rangle = 0.005$. For each system, a shear measurement was obtained by fitting the observed velocity field with a linearly sheared model of perfect circular rotation. The mean measured shear is $\langle \gamma^{obs} \rangle = 0.020 \pm 0.008$, in agreement with our expectations.

Compared to our predicted shears, the distribution of observed shears is broader than can be explained by errors alone, which motivated the main aim of this paper: describing and quantifying different possible sources capable of inducing scatter in the observed shear distribution. Apart from observational uncertainties, in Section \ref{sec:analysis} we have described the two other main sources of scatter, an effective shape noise and a dispersion in the SHMR. We discuss how deviations from axisymmetry in the velocity fields (the underlying assumption of PWL) propagate through shear measurements with a $\sin(2\phi)$ dependence, which ensures that the effective shape noise must have a symmetric distribution centred at zero. We also argue that physical differences in the properties of lenses in the form of a dispersion in the SHMR will result in a skewed distribution in shears. Following the literature, we assumed this distribution to be log-normal.

For each galaxy, we have generated a model for the distribution of possible observed shears as a log-normal dispersion distribution with median $\gamma_{pred}$ convoluted with: 1) a Gaussian effective shape noise modulated by $\sin(2\phi)$ and 2) a Gaussian measurement error. We constructed a likelihood estimator for the parameters $\xi_\gamma$ and $\sigma_\gamma$ representing the dispersion in shears and the effective shape noise. In Section \ref{sec:results}, we present our maximum likelihood measurements of $\sigma_\gamma = 0.024 \pm 0.007$ and $\xi_\gamma = 0.53^{+0.26}_{-0.28}$\,dex.

In section \ref{sec:results} we discuss the importance of a method to self-calibrate PWL samples against an effective shape noise and highlight the low impact of shape noise in our sample, which is $10$ times less affected by noise than conventional WL. We also show that, even with a sample of only 21 PWL measurements, our mean shear measurement is currently limited by data quality and not shape noise (see Section \ref{sec:4.1}). Finally, we quantify the lognormal dispersion in shears due to astrophysical variations in the properties of the lenses: $\xi_\gamma = 0.53^{+0.26}_{-0.28}$\,dex. While the uncertainties are large, this value would be high compared to naive expectations based on models and measurements of the SHMR. Without speculating, we discuss several possible explanations for this suggestive result, including a larger dispersion around the SHMR, variations in inner halo profiles, and/or halo substructure.

The overall result of this paper has been to provide a way to self-calibrate PWL experiments and demonstrate the potential of these techniques given the low impact of shape noise. However, at the moment this unprecedented precision comes at the cost of rather expensive observations and dedicated target selection. As PWL needs well-resolved velocity maps, observations are costly and science targets are limited to bright and nearby galaxies. This cost motivates the use of PWL to analyse rare and high-value systems where the effects of lensing are more apparent or where it is not feasible to build signal by increasing sample size.  In turn, this `sweet spot' requires that targets have been selected in advance, and with high-value systems often being at very low impact parameters, PWL needs more/better spectroscopic surveys with good pair completeness \citep[see for example][]{ deBurghDay15b}. In the future, we expect surveys like DESI-BGS \citep{DESI16}, WAVES \citep{Driver19} and especially a proposed 4MOST Hemisphere Survey to provide many new candidates for PWL. At the same time, with telescopes like the Square Kilometer Array routinely outputting thousands of well-resolved velocity fields, we expect PWL to provide a new avenue for larger cosmology experiments.

\section*{Data availability}
The data underlying this article are available in the gSTAR Data Management and Collaboration Platform at \url{http://dx.doi.org/10.26185/5f488683e4867}.

\section*{Acknowledgements}
We thank the referee for all their valuable input and careful analysis of our work.
This research is partially funded by the Australian Government through an Australian Research Council Future Fellowship (FT150100269) awarded to ENT.
This research made use of Astropy, \footnote{ http://www.astropy.org} a community-developed core Python package for Astronomy, \citep{astropy:2013, astropy:2018}, NumPy \citep{numpy} and CMasher \citep[][]{Ellert2020}.

\bibliographystyle{mnras}
\bibliography{library.bib}

\begin{thebibliography}{}
\makeatletter
\relax
\def\mn@urlcharsother{\let\do\@makeother \do\$\do\&\do\#\do\^\do\_\do\%\do\~}
\def\mn@doi{\begingroup\mn@urlcharsother \@ifnextchar [ {\mn@doi@}
  {\mn@doi@[]}}
\def\mn@doi@[#1]#2{\def\@tempa{#1}\ifx\@tempa\@empty \href
  {http://dx.doi.org/#2} {doi:#2}\else \href {http://dx.doi.org/#2} {#1}\fi
  \endgroup}
\def\mn@eprint#1#2{\mn@eprint@#1:#2::\@nil}
\def\mn@eprint@arXiv#1{\href {http://arxiv.org/abs/#1} {{\tt arXiv:#1}}}
\def\mn@eprint@dblp#1{\href {http://dblp.uni-trier.de/rec/bibtex/#1.xml}
  {dblp:#1}}
\def\mn@eprint@#1:#2:#3:#4\@nil{\def\@tempa {#1}\def\@tempb {#2}\def\@tempc
  {#3}\ifx \@tempc \@empty \let \@tempc \@tempb \let \@tempb \@tempa \fi \ifx
  \@tempb \@empty \def\@tempb {arXiv}\fi \@ifundefined
  {mn@eprint@\@tempb}{\@tempb:\@tempc}{\expandafter \expandafter \csname
  mn@eprint@\@tempb\endcsname \expandafter{\@tempc}}}

\bibitem[\protect\citeauthoryear{{Aihara} et~al.,}{{Aihara}
  et~al.}{2011}]{Aihara11}
{Aihara} H.,  et~al., 2011, \mn@doi [\apjs] {10.1088/0067-0049/193/2/29}, \href
  {https://ui.adsabs.harvard.edu/abs/2011ApJS..193...29A} {193, 29}

\bibitem[\protect\citeauthoryear{{Astropy Collaboration} et~al.,}{{Astropy
  Collaboration} et~al.}{2013}]{astropy:2013}
{Astropy Collaboration} et~al., 2013, \mn@doi [\aap]
  {10.1051/0004-6361/201322068}, \href
  {http://adsabs.harvard.edu/abs/2013A%26A...558A..33A} {558, A33}

\bibitem[\protect\citeauthoryear{{Bartelmann} \& {Schneider}}{{Bartelmann} \&
  {Schneider}}{2001}]{Bartelmann01}
{Bartelmann} M.,  {Schneider} P.,  2001, \mn@doi [\physrep]
  {10.1016/S0370-1573(00)00082-X}, \href
  {https://ui.adsabs.harvard.edu/abs/2001PhR...340..291B} {340, 291}

\bibitem[\protect\citeauthoryear{{Behroozi}, {Conroy}  \&
  {Wechsler}}{{Behroozi} et~al.}{2010}]{Behroozi10}
{Behroozi} P.~S.,  {Conroy} C.,   {Wechsler} R.~H.,  2010, \mn@doi [\apj]
  {10.1088/0004-637X/717/1/379}, \href
  {http://adsabs.harvard.edu/abs/2010ApJ...717..379B} {717, 379}

\bibitem[\protect\citeauthoryear{{Blain}}{{Blain}}{2002}]{Blain02}
{Blain} A.~W.,  2002, \mn@doi [\apjl] {10.1086/341103}, \href
  {https://ui.adsabs.harvard.edu/abs/2002ApJ...570L..51B} {570, L51}

\bibitem[\protect\citeauthoryear{Brown \& Battye}{Brown \&
  Battye}{2011}]{Brown11}
Brown M.~L.,  Battye R.~A.,  2011, \mn@doi [\apj]
  {10.1088/2041-8205/735/1/l23}, 735, L23

\bibitem[\protect\citeauthoryear{{Chambers} et~al.,}{{Chambers}
  et~al.}{2016}]{Chambers16}
{Chambers} K.~C.,  et~al., 2016, arXiv e-prints, \href
  {https://ui.adsabs.harvard.edu/abs/2016arXiv161205560C} {p. arXiv:1612.05560}

\bibitem[\protect\citeauthoryear{{Colless} et~al.,}{{Colless}
  et~al.}{2001}]{Colless01}
{Colless} M.,  et~al., 2001, \mn@doi [\mnras]
  {10.1046/j.1365-8711.2001.04902.x}, \href
  {https://ui.adsabs.harvard.edu/abs/2001MNRAS.328.1039C} {328, 1039}

\bibitem[\protect\citeauthoryear{{Croft}, {Freeman}, {Schuster}  \&
  {Schafer}}{{Croft} et~al.}{2017}]{Croft17}
{Croft} R. A.~C.,  {Freeman} P.~E.,  {Schuster} T.~S.,   {Schafer} C.~M.,
  2017, \mn@doi [\mnras] {10.1093/mnras/stx1206}, \href
  {https://ui.adsabs.harvard.edu/abs/2017MNRAS.469.4422C} {469, 4422}

\bibitem[\protect\citeauthoryear{{DESI Collaboration} et~al.,}{{DESI
  Collaboration} et~al.}{2016}]{DESI16}
{DESI Collaboration} et~al., 2016, arXiv e-prints, \href
  {https://ui.adsabs.harvard.edu/abs/2016arXiv161100036D} {p. arXiv:1611.00036}

\bibitem[\protect\citeauthoryear{{Dopita}, {Hart}, {McGregor}, {Oates},
  {Bloxham}  \& {Jones}}{{Dopita} et~al.}{2007}]{Dopita07}
{Dopita} M.,  {Hart} J.,  {McGregor} P.,  {Oates} P.,  {Bloxham} G.,   {Jones}
  D.,  2007, \mn@doi [\apss] {10.1007/s10509-007-9510-z}, \href
  {https://ui.adsabs.harvard.edu/abs/2007Ap&SS.310..255D} {310, 255}

\bibitem[\protect\citeauthoryear{{Dopita} et~al.,}{{Dopita}
  et~al.}{2010}]{Dopita10}
{Dopita} M.,  et~al., 2010, \mn@doi [\apss] {10.1007/s10509-010-0335-9}, \href
  {https://ui.adsabs.harvard.edu/abs/2010Ap&SS.327..245D} {327, 245}

\bibitem[\protect\citeauthoryear{{Driver} et~al.,}{{Driver}
  et~al.}{2011}]{Driver11}
{Driver} S.~P.,  et~al., 2011, \mn@doi [\mnras]
  {10.1111/j.1365-2966.2010.18188.x}, \href
  {https://ui.adsabs.harvard.edu/abs/2011MNRAS.413..971D} {413, 971}

\bibitem[\protect\citeauthoryear{{Driver} et~al.,}{{Driver}
  et~al.}{2019}]{Driver19}
{Driver} S.~P.,  et~al., 2019, \mn@doi [The Messenger]
  {10.18727/0722-6691/5126}, \href
  {https://ui.adsabs.harvard.edu/abs/2019Msngr.175...46D} {175, 46}

\bibitem[\protect\citeauthoryear{{Duffy}, {Schaye}, {Kay}  \& {Dalla
  Vecchia}}{{Duffy} et~al.}{2008}]{Duffy08}
{Duffy} A.~R.,  {Schaye} J.,  {Kay} S.~T.,   {Dalla Vecchia} C.,  2008, \mn@doi
  [\mnras] {10.1111/j.1745-3933.2008.00537.x}, \href
  {https://ui.adsabs.harvard.edu/abs/2008MNRAS.390L..64D} {390, L64}

\bibitem[\protect\citeauthoryear{{Dvornik} et~al.,}{{Dvornik}
  et~al.}{2020}]{Dvornik20}
{Dvornik} A.,  et~al., 2020, \mn@doi [\aap] {10.1051/0004-6361/202038693},
  \href {https://ui.adsabs.harvard.edu/abs/2020A&A...642A..83D} {642, A83}

\bibitem[\protect\citeauthoryear{{Gurri}, {Taylor}  \& {Fluke}}{{Gurri}
  et~al.}{2020}]{Gurri20}
{Gurri} P.,  {Taylor} E.~N.,   {Fluke} C.~J.,  2020, \mn@doi [\mnras]
  {10.1093/mnras/staa2893}, \href
  {https://ui.adsabs.harvard.edu/abs/2020MNRAS.499.4591G} {499, 4591}

\bibitem[\protect\citeauthoryear{Harris et~al.,}{Harris et~al.}{2020}]{numpy}
Harris C.~R.,  et~al., 2020, \mn@doi [Nature] {10.1038/s41586-020-2649-2}, 585,
  357

\bibitem[\protect\citeauthoryear{{Hoekstra}}{{Hoekstra}}{2013}]{Hoekstra13}
{Hoekstra} H.,  2013, arXiv e-prints, \href
  {https://ui.adsabs.harvard.edu/abs/2013arXiv1312.5981H} {p. arXiv:1312.5981}

\bibitem[\protect\citeauthoryear{{Hoekstra} \& {Jain}}{{Hoekstra} \&
  {Jain}}{2008}]{Hoekstra08}
{Hoekstra} H.,  {Jain} B.,  2008, \mn@doi [Annual Review of Nuclear and
  Particle Science] {10.1146/annurev.nucl.58.110707.171151}, \href
  {https://ui.adsabs.harvard.edu/abs/2008ARNPS..58...99H} {58, 99}

\bibitem[\protect\citeauthoryear{{Huff}, {Krause}, {Eifler}, {Fang}, {George}
  \& {Schlegel}}{{Huff} et~al.}{2013}]{Huff13}
{Huff} E.~M.,  {Krause} E.,  {Eifler} T.,  {Fang} X.,  {George} M.~R.,
  {Schlegel} D.,  2013, arXiv e-prints, \href
  {https://ui.adsabs.harvard.edu/abs/2013arXiv1311.1489H} {p. arXiv:1311.1489}

\bibitem[\protect\citeauthoryear{{Jones} et~al.,}{{Jones}
  et~al.}{2009}]{Jones09}
{Jones} D.~H.,  et~al., 2009, \mn@doi [\mnras]
  {10.1111/j.1365-2966.2009.15338.x}, \href
  {https://ui.adsabs.harvard.edu/abs/2009MNRAS.399..683J} {399, 683}

\bibitem[\protect\citeauthoryear{{Kuijken} et~al.,}{{Kuijken}
  et~al.}{2015}]{Kuijken15}
{Kuijken} K.,  et~al., 2015, \mn@doi [\mnras] {10.1093/mnras/stv2140}, \href
  {https://ui.adsabs.harvard.edu/abs/2015MNRAS.454.3500K} {454, 3500}

\bibitem[\protect\citeauthoryear{{Lagos}, {Tobar}, {Robotham}, {Obreschkow},
  {Mitchell}, {Power}  \& {Elahi}}{{Lagos} et~al.}{2018}]{Lagos18}
{Lagos} C. d.~P.,  {Tobar} R.~J.,  {Robotham} A. S.~G.,  {Obreschkow} D.,
  {Mitchell} P.~D.,  {Power} C.,   {Elahi} P.~J.,  2018, \mn@doi [\mnras]
  {10.1093/mnras/sty2440}, \href
  {https://ui.adsabs.harvard.edu/abs/2018MNRAS.481.3573L} {481, 3573}

\bibitem[\protect\citeauthoryear{{Lange}, {van den Bosch}, {Zentner}, {Wang}
  \& {Villarreal}}{{Lange} et~al.}{2019}]{Lange19}
{Lange} J.~U.,  {van den Bosch} F.~C.,  {Zentner} A.~R.,  {Wang} K.,
  {Villarreal} A.~S.,  2019, \mn@doi [\mnras] {10.1093/mnras/stz1466}, \href
  {https://ui.adsabs.harvard.edu/abs/2019MNRAS.487.3112L} {487, 3112}

\bibitem[\protect\citeauthoryear{{Leauthaud} et~al.,}{{Leauthaud}
  et~al.}{2007}]{Leauthaud07}
{Leauthaud} A.,  et~al., 2007, \mn@doi [\apjs] {10.1086/516598}, \href
  {https://ui.adsabs.harvard.edu/abs/2007ApJS..172..219L} {172, 219}

\bibitem[\protect\citeauthoryear{{Li}, {Wang}  \& {Jing}}{{Li}
  et~al.}{2013}]{Li13}
{Li} C.,  {Wang} L.,   {Jing} Y.~P.,  2013, \mn@doi [\apjl]
  {10.1088/2041-8205/762/1/L7}, \href
  {https://ui.adsabs.harvard.edu/abs/2013ApJ...762L...7L} {762, L7}

\bibitem[\protect\citeauthoryear{{Liske} et~al.,}{{Liske}
  et~al.}{2015}]{Liske15}
{Liske} J.,  et~al., 2015, \mn@doi [\mnras] {10.1093/mnras/stv1436}, \href
  {https://ui.adsabs.harvard.edu/abs/2015MNRAS.452.2087L} {452, 2087}

\bibitem[\protect\citeauthoryear{{Mandelbaum}, {Seljak}, {Kauffmann}, {Hirata}
  \& {Brinkmann}}{{Mandelbaum} et~al.}{2006}]{Mandelbaum06}
{Mandelbaum} R.,  {Seljak} U.,  {Kauffmann} G.,  {Hirata} C.~M.,   {Brinkmann}
  J.,  2006, \mn@doi [\mnras] {10.1111/j.1365-2966.2006.10156.x}, \href
  {https://ui.adsabs.harvard.edu/abs/2006MNRAS.368..715M} {368, 715}

\bibitem[\protect\citeauthoryear{{Miralda-Escude}}{{Miralda-Escude}}{1991}]{Miralda-Escude91a}
{Miralda-Escude} J.,  1991, \mn@doi [\apj] {10.1086/169789}, \href
  {https://ui.adsabs.harvard.edu/abs/1991ApJ...370....1M} {370, 1}

\bibitem[\protect\citeauthoryear{{Mo}, {van den Bosch}  \& {White}}{{Mo}
  et~al.}{2010}]{Mo10}
{Mo} H.,  {van den Bosch} F.~C.,   {White} S.,  2010, {Galaxy Formation and
  Evolution}

\bibitem[\protect\citeauthoryear{{Morales}}{{Morales}}{2006}]{Morales06}
{Morales} M.~F.,  2006, \mn@doi [\apjl] {10.1086/508614}, \href
  {https://ui.adsabs.harvard.edu/abs/2006ApJ...650L..21M} {650, L21}

\bibitem[\protect\citeauthoryear{{Moster}, {Somerville}, {Maulbetsch}, {van den
  Bosch}, {Macci{\`o}}, {Naab}  \& {Oser}}{{Moster} et~al.}{2010}]{Moster10}
{Moster} B.~P.,  {Somerville} R.~S.,  {Maulbetsch} C.,  {van den Bosch} F.~C.,
  {Macci{\`o}} A.~V.,  {Naab} T.,   {Oser} L.,  2010, \mn@doi [\apj]
  {10.1088/0004-637X/710/2/903}, \href
  {https://ui.adsabs.harvard.edu/abs/2010ApJ...710..903M} {710, 903}

\bibitem[\protect\citeauthoryear{{Moster}, {Naab}  \& {White}}{{Moster}
  et~al.}{2013}]{Moster13}
{Moster} B.~P.,  {Naab} T.,   {White} S. D.~M.,  2013, \mn@doi [\mnras]
  {10.1093/mnras/sts261}, \href
  {https://ui.adsabs.harvard.edu/abs/2013MNRAS.428.3121M} {428, 3121}

\bibitem[\protect\citeauthoryear{{Navarro}, {Frenk}  \& {White}}{{Navarro}
  et~al.}{1996}]{Navarro96}
{Navarro} J.~F.,  {Frenk} C.~S.,   {White} S.~D.~M.,  1996, \mn@doi [\apj]
  {10.1086/177173}, \href {http://adsabs.harvard.edu/abs/1996ApJ...462..563N}
  {462, 563}

\bibitem[\protect\citeauthoryear{{Niemi}, {Kitching}  \& {Cropper}}{{Niemi}
  et~al.}{2015}]{Niemi15}
{Niemi} S.-M.,  {Kitching} T.~D.,   {Cropper} M.,  2015, \mn@doi [\mnras]
  {10.1093/mnras/stv2059}, \href
  {https://ui.adsabs.harvard.edu/abs/2015MNRAS.454.1221N} {454, 1221}

\bibitem[\protect\citeauthoryear{{Pillepich} et~al.,}{{Pillepich}
  et~al.}{2018}]{Pillepich18}
{Pillepich} A.,  et~al., 2018, \mn@doi [\mnras] {10.1093/mnras/stx2656}, \href
  {https://ui.adsabs.harvard.edu/abs/2018MNRAS.473.4077P} {473, 4077}

\bibitem[\protect\citeauthoryear{{Price-Whelan} et~al.,}{{Price-Whelan}
  et~al.}{2018}]{astropy:2018}
{Price-Whelan} A.~M.,  et~al., 2018, \mn@doi [\aj] {10.3847/1538-3881/aabc4f},
  \href {https://ui.adsabs.harvard.edu/#abs/2018AJ....156..123T} {156, 123}

\bibitem[\protect\citeauthoryear{{Reddick}, {Wechsler}, {Tinker}  \&
  {Behroozi}}{{Reddick} et~al.}{2013}]{Reddick13}
{Reddick} R.~M.,  {Wechsler} R.~H.,  {Tinker} J.~L.,   {Behroozi} P.~S.,  2013,
  \mn@doi [\apj] {10.1088/0004-637X/771/1/30}, \href
  {https://ui.adsabs.harvard.edu/abs/2013ApJ...771...30R} {771, 30}

\bibitem[\protect\citeauthoryear{{Rodr{\'\i}guez-Puebla}, {Avila-Reese},
  {Yang}, {Foucaud}, {Drory}  \& {Jing}}{{Rodr{\'\i}guez-Puebla}
  et~al.}{2015}]{RPuebla15}
{Rodr{\'\i}guez-Puebla} A.,  {Avila-Reese} V.,  {Yang} X.,  {Foucaud} S.,
  {Drory} N.,   {Jing} Y.~P.,  2015, \mn@doi [\apj]
  {10.1088/0004-637X/799/2/130}, \href
  {https://ui.adsabs.harvard.edu/abs/2015ApJ...799..130R} {799, 130}

\bibitem[\protect\citeauthoryear{{S{\'a}nchez} et~al.,}{{S{\'a}nchez}
  et~al.}{2012}]{Sanchez10}
{S{\'a}nchez} S.~F.,  et~al., 2012, \mn@doi [\aap]
  {10.1051/0004-6361/201117353}, \href
  {https://ui.adsabs.harvard.edu/abs/2012A%26A...538A...8S} {538, A8}

\bibitem[\protect\citeauthoryear{{Sif{\'o}n} et~al.,}{{Sif{\'o}n}
  et~al.}{2015}]{Sifon15}
{Sif{\'o}n} C.,  et~al., 2015, \mn@doi [\mnras] {10.1093/mnras/stv2051}, \href
  {https://ui.adsabs.harvard.edu/abs/2015MNRAS.454.3938S} {454, 3938}

\bibitem[\protect\citeauthoryear{{Springer}, {Ofek}, {Weiss}  \&
  {Merten}}{{Springer} et~al.}{2020}]{Springer20}
{Springer} O.~M.,  {Ofek} E.~O.,  {Weiss} Y.,   {Merten} J.,  2020, \mn@doi
  [\mnras] {10.1093/mnras/stz2991}, \href
  {https://ui.adsabs.harvard.edu/abs/2020MNRAS.491.5301S} {491, 5301}

\bibitem[\protect\citeauthoryear{{Stevens}, {Croton}  \& {Mutch}}{{Stevens}
  et~al.}{2016}]{Stevens16}
{Stevens} A. R.~H.,  {Croton} D.~J.,   {Mutch} S.~J.,  2016, \mn@doi [\mnras]
  {10.1093/mnras/stw1332}, \href
  {https://ui.adsabs.harvard.edu/abs/2016MNRAS.461..859S} {461, 859}

\bibitem[\protect\citeauthoryear{{Taylor} et~al.,}{{Taylor}
  et~al.}{2011}]{Taylor11}
{Taylor} E.~N.,  et~al., 2011, \mn@doi [\mnras]
  {10.1111/j.1365-2966.2011.19536.x}, \href
  {https://ui.adsabs.harvard.edu/abs/2011MNRAS.418.1587T} {418, 1587}

\bibitem[\protect\citeauthoryear{{Taylor} et~al.,}{{Taylor}
  et~al.}{2020}]{Taylor20}
{Taylor} E.~N.,  et~al., 2020, \mn@doi [\mnras] {10.1093/mnras/staa2648}, \href
  {https://ui.adsabs.harvard.edu/abs/2020MNRAS.499.2896T} {499, 2896}

\bibitem[\protect\citeauthoryear{{Viola} et~al.,}{{Viola}
  et~al.}{2015}]{Viola15}
{Viola} M.,  et~al., 2015, \mn@doi [\mnras] {10.1093/mnras/stv1447}, \href
  {https://ui.adsabs.harvard.edu/abs/2015MNRAS.452.3529V} {452, 3529}

\bibitem[\protect\citeauthoryear{{Wright} \& {Brainerd}}{{Wright} \&
  {Brainerd}}{2000}]{Wright00}
{Wright} C.~O.,  {Brainerd} T.~G.,  2000, \mn@doi [\apj] {10.1086/308744},
  \href {https://ui.adsabs.harvard.edu/abs/2000ApJ...534...34W} {534, 34}

\bibitem[\protect\citeauthoryear{{Wright} et~al.,}{{Wright}
  et~al.}{2016}]{Wright16}
{Wright} A.~H.,  et~al., 2016, \mn@doi [\mnras] {10.1093/mnras/stw832}, \href
  {https://ui.adsabs.harvard.edu/abs/2016MNRAS.460..765W} {460, 765}

\bibitem[\protect\citeauthoryear{{Zu} \& {Mandelbaum}}{{Zu} \&
  {Mandelbaum}}{2015}]{Zu15}
{Zu} Y.,  {Mandelbaum} R.,  2015, \mn@doi [\mnras] {10.1093/mnras/stv2062},
  \href {https://ui.adsabs.harvard.edu/abs/2015MNRAS.454.1161Z} {454, 1161}

\bibitem[\protect\citeauthoryear{{de Burgh-Day}, {Taylor}, {Webster}  \&
  {Hopkins}}{{de Burgh-Day} et~al.}{2015}]{deBurghDay15}
{de Burgh-Day} C.~O.,  {Taylor} E.~N.,  {Webster} R.~L.,   {Hopkins} A.~M.,
  2015, \mn@doi [\mnras] {10.1093/mnras/stv1083}, \href
  {https://ui.adsabs.harvard.edu/abs/2015MNRAS.451.2161D} {451, 2161}

\bibitem[\protect\citeauthoryear{{de Burgh-Day}, {Taylor}, {Webster}  \&
  {Hopkins}}{{de Burgh-Day} et~al.}{2016}]{deBurghDay15b}
{de Burgh-Day} C.~O.,  {Taylor} E.~N.,  {Webster} R.~L.,   {Hopkins} A.~M.,
  2016, \mn@doi [\pasa] {10.1017/pasa.2015.39}, \href
  {https://ui.adsabs.harvard.edu/abs/2015PASA...32...40D} {32, e040}

\bibitem[\protect\citeauthoryear{{van Uitert} et~al.,}{{van Uitert}
  et~al.}{2016}]{vanUitert16}
{van Uitert} E.,  et~al., 2016, \mn@doi [\mnras] {10.1093/mnras/stw747}, \href
  {https://ui.adsabs.harvard.edu/abs/2016MNRAS.459.3251V} {459, 3251}

\bibitem[\protect\citeauthoryear{{van der Velden}}{{van der
  Velden}}{2020}]{Ellert2020}
{van der Velden} E.,  2020, \mn@doi [The Journal of Open Source Software]
  {10.21105/joss.02004}, \href
  {https://ui.adsabs.harvard.edu/abs/2020JOSS....5.2004V} {5, 2004}

\makeatother
\end{thebibliography}

\end{document}